\documentclass[%
 reprint,
superscriptaddress,
showpacs,preprintnumbers,
 amsmath,amssymb,
 aps,
pra,
]{revtex4-1}
\usepackage{graphicx}
\usepackage{dcolumn}
\usepackage{bm}
\usepackage[dvipsnames]{xcolor}
\usepackage[colorlinks=true,bookmarks=false,linkcolor=NavyBlue,urlcolor=NavyBlue,citecolor=NavyBlue]{hyperref}
\usepackage{color}

\begin{document}

\title{The role of exchange interaction in nitrogen vacancy centre-based magnetometry}

\author{Cong Son Ho}
\email{elehcs@nus.edu.sg}
 \affiliation{
Department of Electrical and Computer Engineering, National University of Singapore, 4 Engineering Drive 3, Singapore 117576.
}
\author{Seng Ghee Tan}
\affiliation{
Department of Electrical and Computer Engineering, National University of Singapore, 4 Engineering Drive 3, Singapore 117576.
}
\affiliation{
Data Storage Institute, Agency for Science, Technology and Research (A*STAR), 2 Fusionopolis Way, 08-01 Innovis, Singapore 138634. 
}
\author{Mansoor B. A. Jalil}%
\affiliation{
Department of Electrical and Computer Engineering, National University of Singapore, 4 Engineering Drive 3, Singapore 117576.
}
\author{Zilong Chen}
\affiliation{
Data Storage Institute, Agency for Science, Technology and Research (A*STAR), 2 Fusionopolis Way, 08-01 Innovis, Singapore 138634.
}
\affiliation{
Division of Physics and Applied Physics, School of Physical and Mathematical Sciences, Nanyang Technological University, Singapore 637371.}
\author{Leonid A. Krivitsky}
\affiliation{
Data Storage Institute, Agency for Science, Technology and Research (A*STAR), 2 Fusionopolis Way, 08-01 Innovis, Singapore 138634.
}
\date{\today}

\begin{abstract}
We propose a multilayer device comprising of a thin-film-based ferromagnetic hetero-structure (FMH) deposited on a diamond layer doped with nitrogen vacancy centers (NVC's). We  find that when the NVC's are in close proximity (1-2 nm) with the FMH, the exchange energy is comparable to, and may even surpass the magnetostatic interaction energy. This calls for the need to consider and utilize both effects in magnetometry based on NVC's in diamond. As the distance between the FMH and NVC is decreased to the sub-nanometer scale, the exponential increase in the exchange energy suggests spintronic applications of NVC beyond magnetometry, such as detection of spin-Hall effect or spin currents. 
\end{abstract}

\pacs{75.30.Et, 76.30.Mi, 75.50.-y}
\maketitle

\section{Introduction}

 Ferromagnetic hetero-structures (FMH) are integral spintronic elements that play a critical role in various spin transport phenomena. FMH have been used as a platform for practical spintronic devices, such as spin-Hall  \cite{Jung:nat12,Wunder:nat09,Wunder:sci10,Seki:nat08,Miha:prl09} and  spin-orbit-based memories  \cite{Jalil:srep14,Kent:nat15,Matsu:nat15}, anomalous Hall, and topological Hall-based sensors  \cite{Ni:ieee16}. In such systems, the spin density can be detected by many methods, such as magneto-optic Kerr effect (MOKE)  \cite{Kato:sci04,Ver:apl14}, ferromagnetic resonance (FMR) \cite{Liu:nat16}, and electrical measurements  \cite{Lou:prl06,Lou:nat07}. In the characterization of the FMH devices, a spatial resolution of the spin distribution up to the order of $\mathrm{\mu m}$  has been achieved by MOKE \cite{Ver:apl14}, and up to sub-$\mathrm{\mu m}$ using magnetic force microscopy (MFM) \cite{mar:apl87}. However, with the advancement of device fabrication at nanoscale, it requires detection and imaging technique that can provide nanoscale resolution as well as high sensitivity.  

Recently, the nitrogen-vacancy centers (NVC's) in diamond have been employed as extremely sensitive vector magnetic field sensors operating at room temperature and capable of nanoscale spatial resolution  \cite{Tay:nat08,Van:natcom15,arai:nat15,ruga:nat15,Devi:nat15,Grin:nat14,Degen:apl08,Manson:prb06,Acosta:mrs13, Hin:PRAp15}. NVC's are atomic-scale defects which are formed either naturally or by implantation, using high-energy nitrogen ion beam \cite{Ofori:prb12}, annealing  \cite{Nay:prl10, Schwa:jap14}, and nitrogen doping \cite{Ohno:prl12}. The NVC spin states are split under the influence of an external magnetic field. Such spin-splitting selectively suppresses optical fluorescence, making it possible to achieve measurements of the magnetic field through optically detected magnetic resonance (ODMR) \cite{Rodin:rep14,Bala:nat08,Taylor:nat08, Jele:prl04,chi:sci06,De:sci10,Ryan:prl10}.  

In conventional magnetometry experiments  \cite{Rodin:rep14}, the distance between the NVC and the detected magnetic moment $\mathbf{m(r)}$ of an electron is large enough so that only the Zeeman effect has to be considered. The Zeeman effect arises from dipole-dipole interaction which is long range by nature. In sensing the magnetic field profile, a higher spatial resolution can be achieved if the sensor is brought closer to the detected system  \cite{Tay:nat08,Degen:apl08,Rodin:rep14}. At the same time, it has been recently shown that the diamond could be reduced to nanometer scale \cite{Tis:acs11, Mo:nat12,Gong:natcom16} and an NVC could be positioned at a distance of $1-2$ nm below the diamond surface \cite{Loretz:apl14,Acosta:prb09,Ofori:prb12}, thus minimizing the separation between the NVC sensor and the target spin. However, when the sensor is positioned very close to the sample, the electron in NVC can interact with the electron in the sample through the exchange mechanism, in addition to the usual dipole-dipole interaction. In contrast to the dipole interaction, the exchange effect is short range because it depends on the overlap of the NVC wavefunction with that of the magnetic moment $\mathbf{m(r)}$. Therefore, it is important to explore the full nature of electron-electron interaction, {\it {i.e.}}, both the dipole coupling and the exchange coupling, when the NVC is placed in close proximity to the FHM. Recently, the exchange interaction between the ferromagnetic tip and the magnetic sample has also been considered in magnetic exchange force microscopy (MExFM) \cite{Kai:nat07,Pie:prl13,Gran:njp14}. However, to the best of our knowledge, the exchange interaction between proximal NVC spins and FMH spins, has not been considered in NV-based magnetometry. This study will thus be highly relevant for advancing NV-based magnetometry towards its ultimate spatial resolution.

The outline of this paper is as follows. In Sec. \ref{sec2} we formulate the theory of the exchange coupling between FMH spin and NVC spin. In Sec. \ref{sec3} we provide numerical calculation of the exchange coupling, and discuss the feasibility of the proposed scheme and analyze its application in magnetometry. Section \ref{sec4} is the conclusion. 
\section{THEORY}\label{sec2}
\begin{figure}
\centering
 \includegraphics[width=0.45\textwidth]{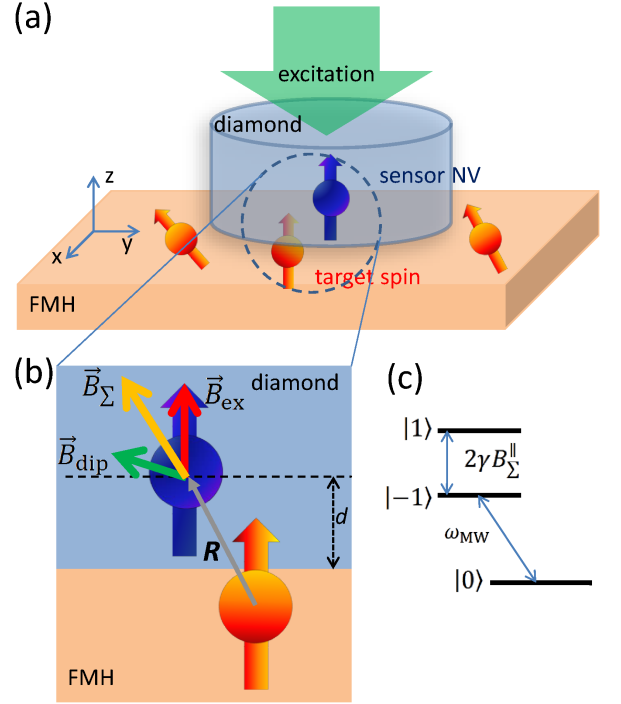}
 \caption{(Color online) (a) Schematic illustration of the physics of FMH magnetic moment interaction with a NVC placed at a short distance above the FMH. (b) The NV spin and FM spin couple via dipole interaction and exchange interaction. Both dipole field $\mathbf{B}_{\mathrm{dip}}$ (green arrow) and exchange field $\mathbf{B}_{\mathrm{ex}}$ (red arrow) are utilized in magnetometry. (c) The Zeeman splitting of the NVC ground state due to the total field $B_\mathrm{\sum}^{\parallel}=B_{\mathrm{dip}}^{\parallel}+B_{\mathrm{ex}}^{\parallel}$, with $(\parallel)$ indicating the field components along the NV axis. The Zeeman splitting is measured via the ODMR technique.
}\label{Fig1}
 \end{figure}

The conceptual device is illustrated in Fig. \ref{Fig1}. We consider a diamond layer engineered such that NVC's are located in close proximity to the diamond surface.  Subsequently, a FMH layer is deposited on top of the diamond surface such that the average distance between the $\mathbf{m(r)}$ of the FMH and the NVC is of the order of a few nanometers \cite{Loretz:apl14,Acosta:prb09,Ofori:prb12}. 

The magnetic dipole field due to the magnetic moment $\mathbf{m}(x,y)$ is given by ${\mathbf{B}}_{\mathrm{dip}}=\frac{{\mu }_0}{4\pi }\left[\frac{3\mathbf{R}\left(\mathbf{m}\cdot \mathbf{R}\right)}{R^5}-\frac{\mathbf{m}}{R^3}\right]$, $\mathbf{R}$ is the displacement from the magnetic moment, and $(x,y)$ are the Cartesian coordinates of the 2D-plane containing the FMH/diamond interface. The Zeeman magnetic energy (dipole coupling) is given by $W_{\mathrm{dip}}={\mathbf{B}}_{\mathrm{dip}}\cdot\ {\mathbf{s}}_{\mathrm{nv}}$, where ${\mathbf{s}}_{\mathrm{nv}}$ is the magnetic moment of the NVC. On the other hand, the exchange energy is given by $W_{\mathrm{ex}}=\mathbf{B}_{\mathrm{ex}}\cdot{\mathbf{s}}_{\mathrm{nv}}$, where $\mathbf{B}_{\mathrm{ex}}=J_{\mathrm{ex}}\left(R\right)\mathbf{m}$ is the exchange field, with the exchange coupling coefficient $J_{\mathrm{ex}}\left(R\right)$ dependent on the distance $R$ between the NVC and the magnetic moment.  Phenomenologically, the distance dependence can be expressed as $J_{\mathrm{ex}}\left(R\right)=J_0\frac{e^{-\gamma R}}{R}$, with $\gamma>0$ being the decay constant. In the following we will analyze in detail the effect of the exchange interaction $W_{\mathrm{ex}}$ on the readout of the NVC response and compare it to the effect of the dipole interaction $W_{\mathrm{dip}}$.
\subsection{ Theory of exchange energy}
The wavefunction for a NVC-FMH system with parallel and anti-parallel spin alignments is given, respectively by: 
\begin{eqnarray}
\psi^{f\sigma,n\sigma}_{1,2\mathrm{P}}&=&\frac{1}{\sqrt{2}}\left[\psi_f(\mathbf{r}_1-\mathbf{R}_i)\psi_n(\mathbf{r}_2)-\psi_f(\mathbf{r}_2-\mathbf{R}_i)\psi_n(\mathbf{r}_1)\right]\nonumber\\
&\times&\left|\xi^{f\sigma}_1\xi^{n\sigma}_2\right\rangle,\\
\psi^{f\sigma',n\sigma}_{1,2\mathrm{AP}}&=&\frac{1}{\sqrt{2}}\left[\psi_f(\mathbf{r}_1-\mathbf{R}_i)\psi_n(\mathbf{r}_2)+\psi_f(\mathbf{r}_2-\mathbf{R}_i)\psi_n(\mathbf{r}_1)\right]\nonumber\\
&\times&\frac{1}{\sqrt{2}}\left|\xi^{f\sigma'}_1\xi^{n\sigma}_2-\xi^{f\sigma'}_2\xi^{n\sigma}_1\right\rangle,
\end{eqnarray}
where $\psi_n(\mathbf{r})$ is the NVC electron wavefunction centered at the vacancy (the origin), and $\psi_f(\mathbf{r}-\mathbf{R}_i )$ is the FMH electron wavefunction located at $\mathbf{R}_i$, and $\xi^{n\sigma}_1 (\xi^{f\sigma}_2)$ represents spin state of electron at the NVC (FMH), respectively, with $\sigma ,\sigma '$ being spin quantum numbers. The parallel wavefunction derives its anti-symmetric property from exchange of position, while the anti-parallel wavefunction derives its antisymmetry from spin exchange. The two-particle wavefunction captures the exchange interaction between NVC and FMH. The interaction strength is determined from the coupling constants:
\begin{eqnarray}
W^\mathrm{P}_{\mathrm{int}}=\int\langle {\psi }^{f\sigma , n\sigma }_{1,2\mathrm{P}}|\mathcal{H}|{\psi }^{f\sigma , n\sigma }_{1,2\mathrm{P}}\rangle d\mathbf{r}_1d\mathbf{r}_2,\\
W^{\mathrm{AP}}_{\mathrm{int}}=\int\langle {\psi }^{f\sigma', n\sigma }_{1,2AP}|\mathcal{H}|{\psi }^{f\sigma' , n\sigma }_{1,2\mathrm{AP}}\rangle d\mathbf{r}_1d\mathbf{r}_2,
\end{eqnarray}
in which $\mathcal{H}=\mathcal{H}_C+\mathcal{H}_B$ is the interaction energy between two particles carrying charge and spin, where $\mathcal{H}_C=\frac{e^2}{4\pi {\epsilon }_0r_{12}}$ is the Coulomb energy, and $\mathcal{H}_B=\frac{3{\mu }_0g^2_e{\mu }^2_B}{8\pi }\left(\frac{1}{r^3_{12}}-\frac{3z^2_{12}}{r^5_{12}}\right)$ is the magnetic dipole interaction energy, with $r_{12}=\left|{\mathbf{r}}_{\mathbf{2}}-{\mathbf{r}}_{\mathbf{1}}\right|, z_{12}=z_1-z_2$. Parallel and anti-parallel configurations (with respect to the spin quantization axis of the NVC) have the same direct coupling energy $W_{\mathrm{dir}}$, while their exchange energy $W_{\mathrm{ex}}$ is opposite in sign. Thus, the exchange physics enables the sensing of orientation of the moment $\mathbf{m}$ in the FMH:
\begin{equation}
W^{\mathbf{P}\mathbf{/}\mathbf{AP}}_{\mathrm{int}}=W_{\mathrm{dir}}\mp W_{\mathrm{ex}} 
\end{equation}
\noindent The above exchange interaction energies modify the unperturbed Hamiltonian of the NVC-FMH system, which can be found, e.g., in  \cite{Doherty:prb12}. Explicitly, we have
\begin{eqnarray}
W^{\mathbf{P}/\mathbf{AP}}_{\mathrm{int}}=\sum^N_i{\int{d{\mathbf{r}}_2}d{\mathbf{r}}_1}|{\psi }_{n}\left({\mathbf{r}}_2\right)|^2\mathcal{H}|{\psi }_f\left({\mathbf{r}}_1\mathbf{-}{\mathbf{R}}_i\right)|^2\nonumber\\
\mp \int{d{\mathbf{r}}_2d{\mathbf{r}}_1}{\psi }^*_{n}\left({\mathbf{r}}_2\right){\psi }^*_f\left({\mathbf{r}}_1\mathbf{-}{\mathbf{R}}_i\right)\mathcal{H}{\psi}_f\left({\mathbf{r}}_2\mathbf{-}{\mathbf{R}}_i\right){\psi }_{n}\left({\mathbf{r}}_1\right).
\end{eqnarray}
\noindent The index $i$ represents the FMH spin center at location $\mathbf{R}_i$, which will be summed over the total number of spin centers in the FMH layer. Physically, the above exchange energy may be expressed as  $W_{\mathrm{ex}}\equiv\sum_i{J_{\mathrm{ex},i}{\mathbf{m}}_{i}\cdot{\mathbf{s}}_{\mathrm{nv}}}$, where the exchange constant $J_{\mathrm{ex},i}$ is a positive scalar with an energy unit (e.g., eV), while ${\mathbf{m}}_{i}\cdot{\mathbf{s}}_{\mathrm{nv}}$ is either $\pm 1$ for P/AP alignment. 
\subsection{Wavefunction of electron spin in FMH}
We assume that a Schottky type barrier is formed at the FMH/diamond interface, with the potential $V_S$ being a function of the distance between the NVC and the interface.  The wavefunction of the FMH penetrates into the diamond, where it exhibits an exponential decay $\psi_f(z)=e^{-\lambda(z)}$, where  $\lambda(z)$ is a decay function and $z$ is the distance from the FMH/diamond interface. Based on the WKB approximation, the decay function can be expressed as 
$\lambda(z)=\int^z_0{\sqrt{\frac{2m_e}{\hbar^2}[V_S(z')-E]}dz'}$,
with $E$ being the electron energy. Taking into account the overlap of the FMH wavefunction within the barrier, the wavefunction of one spin center in the FMH/diamond system can be expressed as follows: 
\begin{eqnarray}
{\psi }_f(\mathbf{r})=\begin{cases} 
      C_1\ F(x,y) g(x,y)e^{-ik_zz} &  (\mathrm{FHM}),\\
      C_2\ F(x,y) g(x,y)e^{-\lambda(z)} & (\mathrm{diamond}). \\
   \end{cases}
\end{eqnarray}
\noindent In the above, $F\left(x,y\right)$ is the in-plane wavefunction, and in the case of free electron, $F\left(x,y\right)=e^{-i(k_xx+k_yy)}$, and $C_1$, $C_2\ $ are the normalization constants. $g(x,y)$ is the distribution function of the FMH wavefunction in the FMH plane, which can have, for example,  a Gaussian form $g(x,y)=\frac{1}{\sqrt{\pi\delta^2}}\exp\left[-\frac{(x-x_0)^2-(y-y_0)^2}{2\delta^2}\right]$, with $(x_0,y_0)$ is the position of FMH spin, and $\delta$ is the width of its wavefunction. For simplicity, the Schottky barrier potential can be approximated as a linear function $V_\mathrm{S}\left(z\right)=\Phi_B\left(1-\frac{\left(\Phi_B-E_c\right)}{\Phi_B}\frac{z}{L_d}\right)$, where $\Phi_B$ is the Schottky barrier height at the FMH/diamond interface, $E_c$ is the bottom of the conduction band of the diamond, and $L_d$ is the depletion length. In this way, the analytical expression of the decaying function and wavefunction can be evaluated, see Appendix \ref{apdix1} for details.
\subsection{ Wavefunction of electron spin in NVC}
\begin{figure}
\centering
 \includegraphics[width=0.5\textwidth]{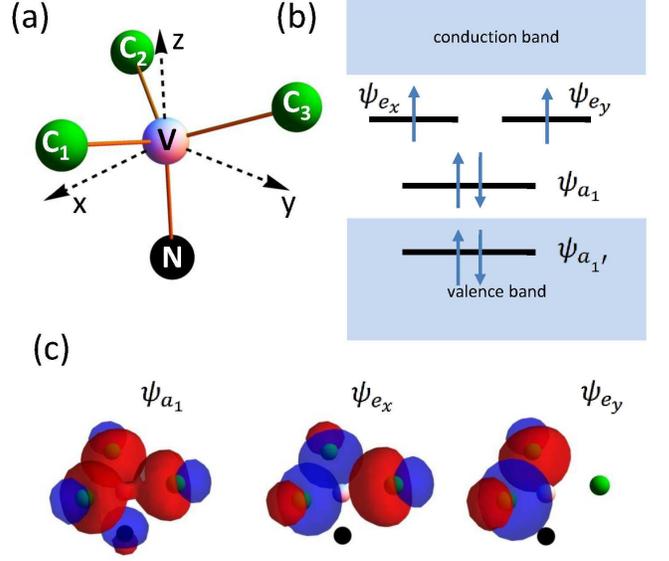}
 \caption{(Color online) (a) Schematic diagram of the tetrahedral structure of the NVC; (b) the energy diagram of the NVC ground states in the absence of the FMH. The energy levels are situated within the bandgap of diamond. (c) The spatial distribution of the electronic orbitals corresponding to the $a_1$, $e_x$ and $e_y$ states.
}\label{Fig2}
 \end{figure}
The NVC, in turn, comprises of wavefunctions ${\phi }_1, {\phi }_2, {\phi }_3$, and $\phi_4$ which can be derived from standard hybridization and normalization approaches. The nitrogen atom $N$ takes on the $sp^3$ configurations along the $z$-direction, while carbon atom $C_3$ takes on the $sp^3$ configurations along the $yz$-plane, see in Fig. \ref{Fig2}. With the vacancy placed at the origin of coordinates, the relative locations of the atoms are fixed. The four $sp^3$ wavefunctions are to be re-parameterized as  ${\phi }^N_1\left(\mathbf{r}-\mathbf{n}\right), {\phi }^{Ci}(\mathbf{r}-{\mathbf{c}}_i)$ to account for a common origin V, see Appendix \ref{SP3}. In the NVC, the individual atomic wavefunctions mix to produce ${\psi }^N_{a_1},\ {\psi }^C_{a_1}\ ,{\psi }_{e_x}, {\psi }_{e_y}$, which are expressed as  \cite{Hoss:prl08}
\begin{eqnarray}
\psi'_{a_1}&=&\phi^N_1,\nonumber\\
\psi_{a_1}&=&\frac{(\phi^{C1}_4+\phi^{C2}_3+\phi^{C3}_2-3S_{nc}\phi^N_1)}{\sqrt{3}\sqrt{1+2S_{cc}-3S^2_{nc}}},\nonumber\\
\psi_{e_x}&=&\frac{(2\phi^{C1}_4-\phi^{C2}_3-\phi^{C3}_2)}{\sqrt{3}\sqrt{2-2S_{cc}}},\label{NV:fun}\\
\psi_{e_y}&=&\frac{(\phi^{C2}_3-\phi^{C3}_2)}{\sqrt{2-2S_{cc}}},\nonumber
\end{eqnarray}
\noindent with $S_{nc}=\left\langle {\phi }^N_1\mathrel{\left|\vphantom{{\phi }^N_1 {\phi }^{C1}_4}\right.\kern-\nulldelimiterspace}{\phi }^{C1}_4\right\rangle $ and $S_{cc}=\left\langle {\phi }^{C1}_4\mathrel{\left|\vphantom{{\phi }^{C1}_4 {\phi }^{C2}_3}\right.\kern-\nulldelimiterspace}{\phi }^{C2}_3\right\rangle$ being the orbital overlap integrals. For these simple wavefunctions $S_{nc}\approx 0.034$ and $S_{cc}\approx 0.067$. Furthermore, the coupling between ${\psi }^N_{a_1}$ and ${\psi }^C_{a_1}$ results in hybridized eigenstates ${\psi }'_{a_1}$ and ${\psi }_{a_1}$. Although there are six electrons in the NVC, two of the electrons occupy state $\psi'_{a_1}$ which lie deep within the valence band \cite{Doherty:prb12}, and hence have negligible contribution to the optical property of the NVC  \cite{Doherty:prb12}. Hence, we need to consider only four electrons, which occupy the states ${\psi }_{a_1}, {\psi }_{e_x},{\psi }_{e_y}$. In bulk diamond, the expanse of the NVC wavefunction is less than 1 nm \cite{Ga:prb08,Hoss:prl08}.
\section{Results and Discussions}\label{sec3}
\subsection{Numerical estimation}
\begin{figure}
\centering
 \includegraphics[width=0.4\textwidth]{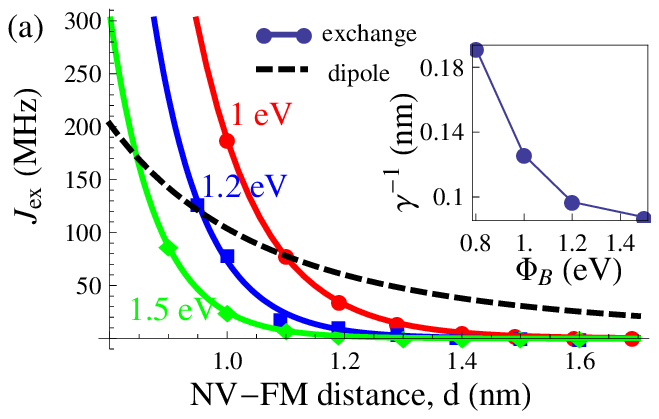}
 \includegraphics[width=0.4\textwidth]{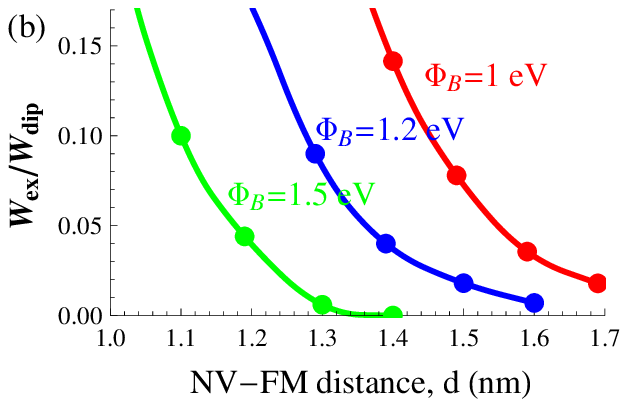}

 \caption{(Color online) (a) Exchange coupling $J_{\mathrm{ex}}$ between electron spin of NVC in $e_x$ state and one FHM spin as a function of NVC-FMH distance $d$ (nm) for various values of the barrier height $\Phi_B$. The numerical calculations are well fitted to the phenomenological formula of the exchange coupling (solid curves). For comparison, the dipole coupling strength is plotted as a dashed curve. Inset: exchange decay constant $\gamma^{-1}$ for various value of $\Phi_B$. (b) Ratio between exchange coupling energy ($W_\mathrm{ex}$) and dipole coupling energy ($W_\mathrm{dip}$). Parameters:  $E_c=E_F, L_d=5\ \mathrm{nm}$, FMH thickness $L_z=1\ \mathrm{nm},\delta=2\ \mathrm{nm}$.
}\label{Fig3}
 \end{figure}
\begin{figure}
\centering
 \includegraphics[width=0.4\textwidth]{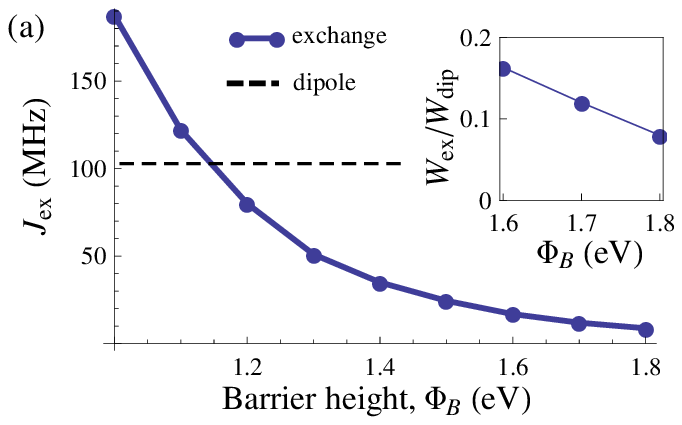}
 \includegraphics[width=0.4\textwidth]{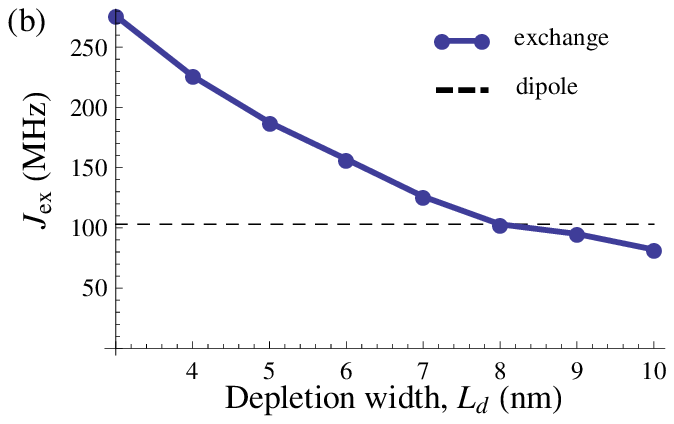}
 \caption{(Color online) Exchange coupling $J_{\mathrm{ex}}$ at a NVC-FMH distance of $d=$1 nm. (a) Exchange coupling as a function of barrier heigth $\Phi_B$, when the depletion length is set at $L_d=5$ nm. Inset: Ratio between exchange coupling energy ($W_\mathrm{ex}$) and dipole coupling energy ($W_\mathrm{dip}$). (b) Exchange coupling as a function of depletion length $L_d$, when the barrier height is set at $\Phi_B=1$ eV.The horizontal dashed lines denote the dipole coupling at $d=1$ nm for comparison. Other parameters:  $E_c=E_F,$ FMH thickness $L_z=1\ \mathrm{nm}$, $\delta=2\ \mathrm{nm}$.
}\label{Fig4}
 \end{figure}
%
 Using the NVC wavefunction given by Eq. \ref{NV:fun}, we can make an estimate of the zero-field ground state splitting, which is essentially a direct exchange between $\left.\left|e_x\right.\right\rangle$ and $\left.\left|e_y\right.\right\rangle$ states, involving the magnetic field energy. The zero-field ground state splitting is given by  \cite{Doherty:prb12}
\begin{eqnarray}
D_{\mathrm{gs}}=\left\langle e_x(\mathbf{r}_1)e_y(\mathbf{r}_2)|\mathcal{H}_B|e_x(\mathbf{r}_1)e_y(\mathbf{r}_2)-e_y(\mathbf{r}_1)e_x(\mathbf{r}_2)\right\rangle,
\end{eqnarray}
\noindent where $\mathcal{H}_B=\frac{3{\mu }_0g^2_e{\mu }^2_B}{8\pi }{\left(\frac{1}{r^3_{12}}-\frac{3z^2_{12}}{r^5_{12}}\right)}$. The estimated ground state splitting is found to be $D_{\mathrm{gs}}\sim 3$ GHz which is close to the measured value of 2.87 GHz  \cite{Acosta:mrs13}. Having affirmed reasonable accuracy of the chosen NVC wavefunction, we consider the NVC-FMH system shown in Fig. \ref{Fig1} and Fig. \ref{Fig2}, with parallel spin alignment represented by ${\psi }^{f\uparrow ,e_x\uparrow}_{1,2\ \mathbf{P}}$ and anti-parallel spin alignment represented by ${\psi }^{f\downarrow,e_x\uparrow }_{1,2\ \mathbf{AP}}$. For simplicity, we consider just the state ${\psi }_{e_x}$ for an NVC deliberately designed to be a spin up pointing vertically to the FMH plane. The NVC wavefunction ${\psi }_{e_x}$ overlaps with ${\psi }_f\left(x,y,z\right)$ emanating from one particular spin center at the FMH. As there are many FMH spin centers, the total effect is summed over the volume of the FMH.

 Results of our calculated exchange coupling between NVC spin and one FMH spin are shown in Fig. \ref{Fig3}. First, it shows that the exchange coupling exponentially decays with respect to the FMH-NVC distance. The phenomenological formula of the exchange coupling $J_{\mathrm{ex}}\propto e^{-\gamma R}/R$ gives a good fit to our numerical results, confirming the sensitivity of the exchange coupling to the distance. Second, a low barrier height $\Phi_B$ and narrow depletion region, which allows deep wavefunction penetration into the diamond, are very important pre-conditions to achieve significant exchange energy, see Fig. \ref{Fig4}.

Meanwhile, as it was shown earlier, the interface Schottky barrier $\Phi_B$ is almost constant regardless of the doping level  \cite{Sun:jap12}, it can be controlled by applying a bias voltage. However, this is limited by the electro-chemical property  of the interface as high voltage may lead to the surface oxidation \cite{Grot:natcom12}. An alternative way to reduce the barrier height at the interface is to terminate the diamond surface with elements that induce a large positive electron affinity (PEA), since the Schottky barrier at the interface is roughly determined by the difference between the work function $(\Phi_M)$ of the metal and the electron affinity of diamond, i.e., $\Phi_B=\Phi_M-\chi_e$. For example, diamond with an oxygen-terminated surface has a PEA value of $\chi_e=2.13$ eV \cite{maier2001electron}, while that with a chlorine-terminated surface has $\chi_e=2.1-2.65$ eV \cite{tiwari2011calculated}, and that with a fluorine-terminated surface has $\chi_e=2.56$ eV \cite{rietwyk2013work}. With the above PEA values and typical work function value of metals $\Phi_M=5$ eV, the barrier can be reduced to $\Phi_B\sim 2.5$ eV. A combination of applied bias and choice of surface termination can result in a significant reduction of the barrier height down to $\sim 1$ eV, which would allow significant penetration of the FMH electron wavefunction into diamond.

 At the same time, the depletion region can be reduced by increasing the doping concentration or by applying a gate voltage across the FMH/diamond interface  \cite{Grot:natcom12, Gei:jvct96,Gei:apl96,Sun:jap12}. In diamond doped with nitrogen ($[N]$), the depletion width is determined by $L_d=\sqrt{\frac{\epsilon_0\epsilon_r (V_{bi}+V_a)}{2\pi e[N]}}$, where $V_{bi}$ is the intrinsic build-in potential \cite{San:prb09, Sun:jap12}, $V_a$ is applied voltage, $\epsilon_r=5.68$ is the dielectric constant of a diamond, and $\epsilon_0$ is the vacuum permittivity. For example, a depletion width of 10 nm can be achieved with an implantation dose of 10 ppm ($1.8\times 10^{18}\ \mathrm{cm}^{-3}$), and it can be reduced to 3 nm for a dose of 100 ppm ($1.8\times 10^{19}\ \mathrm{cm}^{-3}$). As discussed later, this narrow depletion width is also crucial for stability of negatively-charged NV centers, but high implantation dose will degrade the spin-coherence of the NV centers. In general, a wider depletion width will reduce the wavefunction overlap between the NV center and the magnetic moment in the FM layer, and hence the exchange coupling between the two. However, the degree of the overlap (and thus the strength of the exchange coupling) is a weaker function of the depletion width compared to that of the Schottky barrier height (see Appendix \ref{apdix1}). For instance even at a depletion width of 10 nm, the exchange coupling strength is still roughly of the same order of magnitude as that corresponding to a depletion width of a few nm (say 3 nm) - see Fig. \ref{Fig4}(b).

 Given the FMH/NVC distance achievable by today's technology of about 1-2 nm, a barrier height (with respect to the Fermi level of FMH) of less than 1 eV will already result in the exchange energy surpassing the magnetic energy (see Fig. \ref{Fig3} (a)). Under this circumstance, it is essential to consider the NVC-FMH exchange interaction in the design of NVC-based magnetometry devices. Both magnetic and exchange energies are estimated to be at the order of a few hundred MHz of the 2.87 GHz zero field splitting of the NVC. For separation distance more than 2 nm, the exchange energy rapidly decreases to zero, at a much faster rate than the dipole coupling energy ( see Fig.\ref{Fig3}). This is consistent with prevailing experimental focus on using only the magnetic energy in describing the results of the magnetometry experiments.  Considering ongoing progress in development of etching and lithography tools which become increasingly capable of producing hetero-structures with small distances between NVC and the interface, it makes perfect sense to start considering the short-range exchange effect. 

\subsection{Stability of shallow $\mathrm{NV}^-$ centers}

\begin{figure}[t].
\centering
 \includegraphics[width=0.5\textwidth]{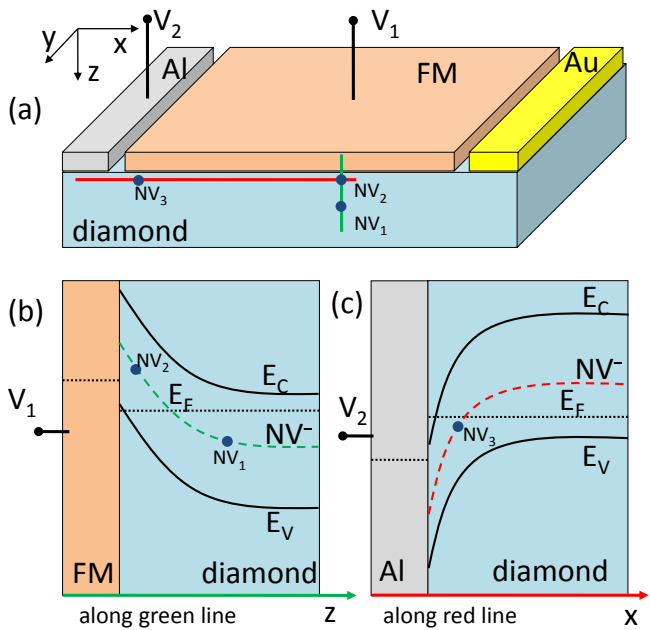}
 \caption{(Color online) Principle of $\mathrm{NV^-}$ stabilization:  (a) Schematic diagram of FM/diamond system with vertical ($V_1$) and in-plane $(V_2$) electrical gates. (b) Energy band diagram across FM/diamond interface (green line). In diamond, the band-bending in the depletion region will raise the $\mathrm{NV}^-$ level above the diamond Fermi level, resulting in discharge of shallow $\mathrm{NV}^-$ to $\mathrm{NV}^0$ ($NV_2$) . (c) Energy band diagram across in-plane diamond surface near Aluminum contact (red line). The cross section is a few nm below the diamond surface. The work function of Al is smaller than that of diamond surface, resulting in a depletion region of holes. An applied reverse voltage will raise the diamond Fermi level above the $\mathrm{NV}^-$. Thus, an NV center near the diamond surface and close to the Al contact can be stabilized in the charged $\mathrm{NV}^-$ state ($\mathrm{NV}_3$).
}\label{Stab}
 \end{figure}

Now we will discuss about the stability of the negatively-charged $\mathrm{NV}^-$ centers near the FMH/diamond interface, which would be a crucial factor for the feasibility of our scheme. The charged state of an NV center depends on the $\mathrm{NV}^{-/0}$ transition level relative to the Fermi level. In diamond doped with nitrogen, the Fermi level is 1.7 eV below the conduction band minimum (CBM) \cite{Collin:2002,farrer1969substitutional} , and the  $\mathrm{NV}^-$ level is 2.58 eV below the CBM \cite{weber2010quantum,steeds2000photoluminescence}. Thus, in free-standing nitrogen-doped diamond, the NV centers deep in the bulk diamond are likely to be in the $\mathrm{NV}^-$ state ($\mathrm{NV}_1$ in Fig. \ref{Stab} (a)). When the diamond with hydrogen-terminated surface is in contact with metal \cite{Gei:jvct96,Gei:apl96,Sun:jap12} or air \cite{Hau:prb11} , part of the $\mathrm{NV}^{-/0}$ level would be raised above the Fermi level due to the band bending near the interface \cite{Hau:prb11,Grot:natcom12,Schrey:scirep14}. As a result, $\mathrm{NV}^-$ (charged state) near the interface has a tendency to be discharged to $\mathrm{NV}^0$ (neutral state) ($NV_2$ in Fig. \ref{Stab}a).

However, the $\mathrm{NV}^-$ charged state  can be  stabilized by controlling the Fermi level and the band bending near the interface via electrical gating \cite{Grot:natcom12,Doi:prx14,Schrey:scirep14,Kara:pnas16}, chemical treatment \cite{Mai:prl00,Collin:2002, Hau:prb11}, or by controlling the charge/discharge process by  optical excitation\cite{Siyu:prl13,chen2013optical,Ji:prb16}, and possibly a combination of the above operations.  For example, diamond with PEA oxygen-terminated surface can stabilize shallow $\mathrm{NV}^-$ centers \cite{fu2010conversion,Hau:prb11}. As discussed in previous section, PEA surfaces can be produced by terminating the surfaces with oxygen, chlorine or fluorine \cite{maier2001electron,tiwari2011calculated,rietwyk2013work}. At the same time, as mentioned early, a high dose of nitrogen (n-dopant) implantation can increase the  band bending and reduce the depletion width \cite{Hau:prb11,Grot:natcom12}, thus shifting the crossing point of the Fermi level and the $\mathrm{NV}^-$ level closer to the interface. Moreover, by applying a gate voltage across the metal/diamond, one can raise Fermi level above the $\mathrm{NV}^{-/0}$ transition level beyond a certain depth \cite{San:prb09, Grot:natcom12}. In a previous work \cite{Grot:natcom12}, the charged $\mathrm{NV}^-$ state is stabilized at a depth of 7 nm from the interface by applying a gate voltage of +0.5 eV, and an implantation dose of $10^{13} \ \mathrm{cm^{-2}}$ \cite{Grot:natcom12}. A higher dose of $10^{14} \ \mathrm{cm^{-2}}$ can even stabilize NV centers at less than 5 nm depth without electrical gating \cite{Hau:prb11}. Moreover, it has been shown that phosphorous doping can also effectively produce a  pure $\mathrm{NV}^-$ population \cite{doi2016pure}. This is based on the fact that phosphorous-donor level is 0.57 eV below CBM \cite{katagiri2004lightly}, which is much lower than that of nitrogen (1.7 eV). In our scheme, we require the charged NV−centers to be only about 2 nm from the interface to induce significant exchange coupling. Thus, a high gate voltage and high implantation dose are required to stabilize the NV−centers so close to the interface. However, there is a limitation to both methods due to possibility of oxidation at high applied voltage and degraded spin-coherence time at high implantation dose \cite{van1997dependences, tallaire2006characterisation,Rodin:rep14,Tay:nat08}. In general, the spin-coherence time is inversely proportional to the concentration of both uncontrolled electron and nuclear spins around the NV center \cite{Rodin:rep14,Tay:nat08}. At high doping level $\mathrm{[N]}>100$ ppm, and with concentration of $\mathrm{^{13}C}$ nuclear spin of 1.1 \%, the coherence-time $T_2^*$ is limited by the $\mathrm{[N]}$ electron spin to $\sim 0.1\ \mu \mathrm{s}(100\mathrm{ppm}/\mathrm{[N]})$. For low doping level $\mathrm{[N]} < 10$ ppm, $T_2^*$ is limited by $\mathrm{^{13}C}$  to $\sim 1\ \mu \mathrm{s}$, and it is can be up to $\sim 100\ \mu \mathrm{s}$ in purified diamond where $\mathrm{[^{13}C]}<0.01\%$ and $\mathrm{[N]}<1$ ppm.

Possible means to overcome these limitations would be to apply the gate voltage in an in-plane configuration  \cite{Schrey:scirep14,shimizu2016charge}, see Fig. \ref{Stab} (b). Using aluminum (Al) as a contact, a lateral hole depletion region is formed near the Al contact with the downward band bending. An applied reverse voltage will raise the diamond Fermi level above the $\mathrm{NV}^-$ level, thus stabilizing shallow $\mathrm{NV}^-$ centers near the Al contact ($\mathrm{NV}_3$ in Fig. \ref{Stab} (b)). 

In addition, one can use optical excitation to convert $\mathrm{NV}^0$ to $\mathrm{NV}^-$ \cite{Siyu:prl13,chen2013optical,Ji:prb16}. The working principle is as follows. Assuming the NV center is in the neutral charge state with unoccupied $e_y$ level (see Fig. \ref{Fig2}), a first laser photon will excite one electron from $a_1$ to $e_y$, followed by another photon that excites another electron from $a_1'$ (deep in valence band) to $a_1$. The net result is the conversion of $\mathrm{NV}^0$ to $\mathrm{NV}^-$ \cite{Siyu:prl13}. Note that to prevent subsequent discharge of $\mathrm{NV}^-$ to $\mathrm{NV}^0$, the laser excitation needs to be applied continuously.

\subsection{Application in magnetometry}
When the NVC spin is coupled to an FMH plane that hosts many spins, the exchange coupling can exist between any pair of NVC-FMH spins. However, as shown in Fig. \ref{Fig3}, the exchange coupling quickly diminishes as the distance increases. Therefore, the exchange coupling between NVC spin and its nearest FMH spin takes the dominant role. Assuming the spin distribution in FMH layer is $\mathbf{m}(x,y)$, and a NVC is placed at position $(x_i,y_i)$, the total exchange energy is then $W_{\mathrm{ex}}\approx \mathbf{B}_{\mathrm{ex}}\cdot \mathbf{s}_{\mathrm{nv}}$, with $B_{\mathrm{ex}}=J_{\mathrm{ex}} m(x_i,y_i)$, which means that the exchange coupling between NVC and FMH is mainly dependent on the local spin density at the NVC position. 

However, the dipole field is significant even at long range, and the Zeeman splitting of the NVC ground state is induced by the combined exchange and dipole field. In contrast to the exchange field, the total dipole field induced by the FMH layer on the NV spin is the sum $B_{\mathrm{dip}}=\sum_{i} B_{\mathrm{dip}}(x_i,y_i)$, where $B_{\mathrm{dip}}(x_i,y_i)$ is the dipole field induced by FMH spin $m(x_i,y_i)$. The dipole field is dependent on the distance $d$ between NVC and FMH layer for a certain NVC orientation. Meanwhile, the exchange coupling $J_{\mathrm{ex}}$ is not only dependent on the NVC-FMH distance $(d)$, but also on properties of the FMH/diamond junction such as barrier potential ($\Phi_B$) and depletion width ($L_d$), see Figs. \ref{Fig3} and \ref{Fig4}. This enables us to tune the exchange coupling and the exchange field by modifying the barrier potential or depletion width, but still keeping the dipole field constant. Based on the above, we will now show that the spin distribution in FMH layer can be read out locally.

First, let us consider a specific configuration of the FMH/NVC system and then perform the ODMR measurement  \cite{Jele:prl04,chi:sci06,De:sci10,Ryan:prl10} to determine the total Zeemann field $B_\mathrm{\sum}^{(0)}=B_{\mathrm{dip}}^{(0)}+B_{\mathrm{ex}}^{(0)}$, where $B_{\mathrm{ex}}^{(0)}= J_{\mathrm{ex}}^{(0)} m(x_i,y_i)$. Let us assume that we can tune one of the junction properties, {\it{i.e.}}, barrier potential $(\Phi_B)$ or depletion width $(L_d)$ while keeping the NVC-FMH distance $(d)$ unchanged. Then, the exchange field is modified to be $B_{\mathrm{ex}}^{(1)}= J_{\mathrm{ex}}^{(1)} m(x_i,y_i)$, while the dipole field is unchanged. The total Zeeman field is now $B_\mathrm{\sum}^{(1)}=B_{\mathrm{dip}}^{(0)}+B_{\mathrm{ex}}^{(1)}$. From the above, we can derive the local spin density at $(x_i,y_i)$ as follows
\begin{equation}\label{eq20}
m(x_i,y_i)= \frac{B_\mathrm{\sum}^{(1)}-B_\mathrm{\sum}^{(0)}}{J_{\mathrm{ex}}^{(1)}-J_{\mathrm{ex}}^{(0)}}.
\end{equation}
Note that Eq. (\ref{eq20}) is obtained by assuming that the exchange coupling is only retained for the most nearest NVC spin- FMH spin pair. However, if the coupling between NVC spin and other nearby FMH spins are taken into account, Eq. (\ref{eq20}) shall represent the average spin density in a region close to the NVC spin, which results in a lower spatial resolution. To determine the region to be averaged over, we assume that the exchange coupling can be significant between NVC spin and FMH spin separated by distance up to $d+\gamma^{-1}$, where $\gamma^{-1}$ is the decay constant (see inset of Fig. \ref{Fig3}). The intersection of the sphere centered at the origin with radius $d+\gamma^{-1}$ and the FMH plane is the region that will be averaged over, which is a circle of diameter $2\sqrt{2d\gamma^{-1}}$. For $d=1-2$ nm and calculated value of $\gamma$ as shown in Fig. \ref{Fig3}, the spatial resolution is estimated to be of the range of 0.8-2 nm. For comparison, the spatial resolution in recent NV-based magnetometry experiment is up to the order of 100 nm  \cite{Hin:PRAp15}.

\section{CONCLUSION}\label{sec4}
We have proposed a novel FMH-NVC system engineered for observation of exchange effects between FMH spins and the NVC spin. We showed that the exchange energy can be substantial, and even surpass the usual magnetic dipole energy for low barrier heights and nanometer separation between the NVC and FMH spins, which is within the capability of present-day nanofabrication technology. This opens up possibilities to characterize FMH spintronic systems with nanometer spatial resolution and high spectroscopic precision using NV magnetometry.

\begin{acknowledgments}
\noindent This work is supported by the Singapore National Research Foundation under Grants: NRF-NRFF2011-07, NRF-CRP14-2014-04, and CRP Programs “Next Generation Spin Torque Memories: From Fundamental Physics to Applications” NRF-CRP12-2013-01 and “Non-Volatile Magnetic Logic and Memory Integrated Circuit Devices” NRF-CRP9-2011-01, as well as the Singapore Ministry of Education Academic Research Fund Tier 1 Project R-263-000-C06-112, and Tier 2 project MOE2013-T2-2-125 (NUS Grant No. R-263-000-B10-112).
\end{acknowledgments}

\appendix

\section{Decaying wavefunction of electron in FMH/diamond structure}\label{apdix1}
The Schottky barrier can be approximated as a linearly decreasing barrier potential
\begin{equation}
V_S\left(z\right)=\Phi_B\left(1-\frac{\left(\Phi_B-E_c\right)}{\Phi_B}\frac{z}{L_d}\right),
\end{equation}
where $\Phi_B$ is the Schottky barrier height at the FMH/diamond interface, $E_c$ is the the conduction band minimum (CBM) of the diamond, $L_d$ is the depletion width. In diamond doped with nitrogen ($N_d$), the depletion width is determined by $L_d=\sqrt{\frac{\epsilon_0\epsilon_r (V_{bi}+V_a)}{2\pi eN_d}}$, where $V_{bi}=2.3$ eV is the intrinsic build-in potential \cite{Sun:jap12}, $V_a$ is applied voltage, $\epsilon_r=5.68$ is the dielectric constant of diamond, and $\epsilon_0$ is the vacuum permittivity. 

\begin{figure}
\centering
 \includegraphics[width=0.5\textwidth]{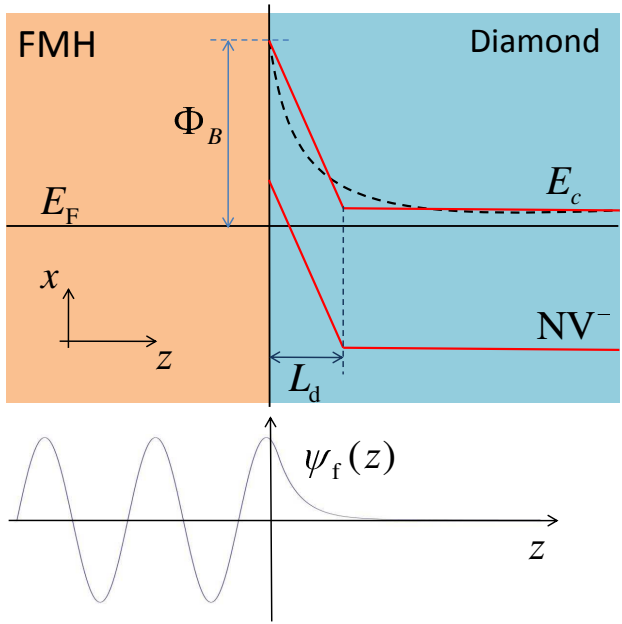}
 \caption{(Color online) Energy-band schematic of FMH/diamond junction. Near the interface, a Schottky-type junction is formed, with Schottky barrier height $\Phi_B$ and depletion width $L_d$ (dashed curve). The Schottky junction is approximately modeled as a linearly decreasing potential (solid red curve). $E_F$ is the Fermi level of the ferromagetic layer, which we take to be zero energy, and $E_c$ is the conduction band minimum of diamond (CBM), $\mathrm{NV}^-$ is the transition level of the negatively-charged state. The wavefunction of electron in FMH has a finite decay length into the diamond (bottom panel). 
}\label{Fig5}
 \end{figure}

Now we consider the transport of electron from FMH to the diamond. The overall wavefunction of electron can be found by solving the Schrodinger equation, which is given by
\begin{equation}
\left(-\frac{{\hbar }^2}{2m_e}\frac{{\partial }^2}{\partial z^2}+\left(U-E\right)\right)\psi \left(z\right)=0,
\end{equation} 
where
\begin{equation}
U=\begin{cases}
0&-L_z<z<0\ (\mathrm{FMH}), \\ 
V_S\left(z\right)& 0<z<L_d\ (\mathrm{diamond}), 
\end{cases}
\end{equation}
in which $L_z$ is the thickness of the FMH.
In FMH region, the electron can be described by wavefunction of free electron:
\begin{equation}
{\psi }_{f,1}\left(z\right)=C_1{\mathrm{sin} \left[k_z\left(z+L_z\right)\right]\ },
\end{equation}
such  that the wavefunction at FMH/vacuum interface is zero, {\it{i.e.}}, ${\psi }_{f,1}\left(-L_z\right)=0$, with $C_1$ is a constant, and $k_z=\sqrt{\frac{2m_eE}{{\hbar }^2}}$. 
In the diamond region, the wavefunction decays as 
\begin{equation}\label{decay}
\psi_{f,2}(z)= C_2 e^{-\lambda(z)},
\end{equation}
where $\lambda(z)$ is decay function of $z$. In WKB formalism, the decay function is expressed as
\begin{eqnarray}
\lambda(z)=\int^z_0{\sqrt{\frac{2m_e}{\hbar^2}[V_S(z')-E]}dz'}.
\end{eqnarray}
Substituting the barrier potential we have:
\begin{eqnarray}
\lambda(z)&&=-\frac{2\sqrt{2m_e}L_d{\left(\Phi_B-E\right)}^{\frac{3}{2}}}{3\hbar (\Phi_B-E_c)}\nonumber\\
+&&\frac{2\sqrt{2m_e}L_d}{3\hbar (\Phi_B-E_c)}{\left(\Phi_B-E-\frac{\left(\Phi_B-E_c\right)z}{L_d}\right)}^{\frac{3}{2}}.
\end{eqnarray} 
The decay wavefunction is illustrated in Fig. \ref{Fig5}.  Note that with the coordinate origin at the vacancy (V), the above wavefunctions are reparameterized as $\psi_f(z+d)$, where $d$ is the FMH/NVC separation distance. The wavefunction penetration is generally enhanced for lower barrier height and smaller depletion width. As shown in previous works \cite{Ga:prb08,Hoss:prl08}, the extent of the of NVC wavefunction is less than 1 nm, which is much less than the depletion width ($L_d \sim 5-10$ nm), i.e., $z\ll L_d$. The decay wavefunction in Eq. \eqref{decay} is approximated as
\begin{equation}\label{decay2}
\psi_{f,2}(z) \sim e^{-\lambda_0 d} e^{-\lambda_0 z (1-\frac{d}{2L_d})},
\end{equation}
where $\lambda_0 =\sqrt{\frac{2m_e\Phi_B}{\hbar^2}}$, and we have assumed $E=E_c=E_F$, with $E_F$ being the FM Fermi energy. From Eq. \eqref{decay2} we can obtain the dependence of the exchange coupling on the FMH/NVC distance and other parameters, i.e., barrier height and depletion width. First, the exchange coupling would exponentially decay as $\sim e^{-\lambda_0 d}$ with increasing $d$, where the decay constant $\lambda_0$ is scaled with the barrier height $\Phi_B$. Second, the depletion width $L_d$ contributes a correction of $d/2L_d$ to the decay function, which is insignificant if $L_d\gg d$. Thus, the strength of the exchange coupling is a weaker function of the depletion width compared to that of the Schottky barrier height.

To obtain the normalization constants, we apply the boundary conditions at the vacuum/FMH interface and FMH/diamond interface, respectively:
\begin{eqnarray}
{\psi }_{f,1}\left(0\right)={\psi }_{f,2}\left(0\right),\nonumber\\
\frac{d}{dz}{\psi}_	{f,1}\left(0\right)=\frac{d}{dz}{\psi }_{f,2}\left(0\right),\label{eqh}
\end{eqnarray}
or
\begin{eqnarray}
\left[ \begin{array}{cc}
h_{11} & h_{12} \\ 
h_{21} & h_{22} \end{array}
\right]\left( \begin{array}{c}
C_1 \\ 
C_2 \end{array}
\right)=0,
\end{eqnarray}
with $h_{ij}$ being the linear coefficients of $C_{1,2}$ deduced from Eq. \ref{eqh}. From this set of equations, we can find the possible energy levels by solving equation ${\mathrm{det} (h)\ }=0$. Once the energy is found, the normalization constants $C_1$ and $C_2$ can be obtained from the normalization condition $\int dz\ \left|{\psi }_f\left(z\right)\right|^2=1$.

\section{$sp^3$ hybridization}\label{SP3}
The hybridized $sp^3$ wavefunctions of the electrons are given by
\begin{gather*}
\phi^N_1=\frac{1}{2}\psi_N^{2s}+\frac{\sqrt{3}}{2}\psi_N^{2p_z},\nonumber\\
\phi^{C3}_2=\frac{1}{2}\psi_C^{2s}+\sqrt{\frac{2}{3}}\psi_C^{2p_y}-\frac{1}{2\sqrt{3}}\psi_C^{2p_z},\nonumber\\
\phi^{C2}_3=\frac{1}{2}\psi_C^{2s}+\frac{1}{\sqrt{2}}\psi_C^{2p_x}-\frac{1}{\sqrt{6}}\psi_C^{2p_y}-\frac{1}{2\sqrt{3}}\psi_C^{2p_z},\nonumber\\
\phi^{C1}_4=\frac{1}{2}\psi_C^{2s}-\frac{1}{\sqrt{2}}\psi_C^{2p_x}-\frac{1}{\sqrt{6}}\psi_C^{2p_y})-\frac{1}{2\sqrt{3}}\psi_C^{2p_z},\nonumber
\end{gather*}
\noindent where $\psi_X^{2s}$ and $\psi_X^{2\mathbf{p}}$ are the wavefunction of 2s and 2p orbital states. If we set the coordinates system so that the origin is at the vacancy (V), the N atom is on the $z$-axis and one of the carbon atom is in $yz-$plane, then $sp^3$ wavefunctions are to be re-parameterized as  ${\phi }^N_1\left(\mathbf{r}-\mathbf{n}\right), {\phi }^{Ci}(\mathbf{r}-\mathbf{c}_i)$, where $\mathbf{n}\approx a \hat{z}$, $\mathbf{c}_1\approx a (0, 0.94, -0.33)$, $\mathbf{c}_2\approx a (0.81, 0.47, 0.33)$, $\mathbf{c}_3\approx a (-0.81, 0.47, 0.33)$ representing the relative position of the nitrogen and carbon atoms, with $a=0.15$ nm being the bond length of the diamond lattice. 
%
%
%
%

%

\end{document}